\documentclass{aa}     
\usepackage{aas_macros}

\usepackage{graphicx}
\usepackage{txfonts}
\usepackage{lipsum}
\usepackage{subcaption}         
\usepackage{lscape}             
\usepackage{placeins}           
\usepackage{hyperref}
\usepackage{color}

\newcommand{\edit}[1]{{#1}}

\begin{document}

   \title{Gravitational scattering of ejecta in the Didymos system cannot explain the evolution of the binary's orbital period}
   \titlerunning{Gravitational scattering of ejecta in the Didymos system}

   \author{Harrison Agrusa\inst{1}
        \and Camille Chatenet\inst{1,2}
        }

   \institute{Universit\'e C\^ote d’Azur, Observatoire de la C\^ote d’Azur, CNRS, Laboratoire Lagrange, Nice, France\\
             \email{hagrusa@oca.eu}
            \and Institut UTINAM, CNRS UMR 6213, Universit\'e Marie et Louis Pasteur, OSU THETA, BP 1615, Besançon, France \\ }
   \date{Received 2 June 2025 / Revised 7 August 2025 / Accepted 24 August 2025}

 
  \abstract
   {In the first few months following the DART impact on Dimorphos, it appears that the orbital period dropped by ${\sim} 30$ s in addition to the immediate ${\sim}30 $ min drop. This effect has been attributed to ``binary hardening,'' whereby the binary's orbital period would have gradually decreased as Dimorphos continuously scattered bound ejecta out of the system and lost angular momentum.}
   {We investigated this hypothesis with the goal of constraining the conditions that would lead to a gradual decrease in the binary's orbital period.} 
   {We used $N$-body simulations to study the dynamical evolution of the Didymos system under the influence of a cloud of massive test particles.}
   {We demonstrate that the gravitational scattering of ejecta is not a plausible explanation for Dimorphos's anomalous orbital period drop under any circumstances. This is a result of Dimorphos's escape speed being low compared to its orbital velocity, making it a weak scatterer.}
   {If a significant fraction of DART ejecta was launched at low speeds, as impact models and scaling laws suggest, then the binary's orbital period was likely increased as this material was accreted back onto Didymos and Dimorphos. Therefore, some additional mechanism must have overcome this effect, leading to a net orbital period decrease.}

   \keywords{ Minor planets, asteroids: individual: (65803) Didymos --
                Celestial mechanics
               }

   \maketitle

\nolinenumbers

\section{Introduction}
In September of 2022, NASA's Double Asteroid Redirection Test (DART) mission successfully collided with Dimorphos, the secondary component of the Didymos--Dimorphos system \citep{Daly2023,Chabot2024}. The impact successfully reduced the binary's orbital period by ${\sim}33$ m \citep{Thomas2023}, owing to significant momentum transfer generated by recoiling ejecta\citep{Cheng2023}. The LICIAcube spacecraft, which was deployed approximately two weeks before impact, managed to capture many images of the Didymos system shortly after the impact, providing an up-close view of the impact ejecta and its unique morphology \citep{Dotto2024}. 

In the weeks and months following the DART impact, a robust post-impact orbit determination of Dimorphos was achieved using two independent methods. Using only ground-based photometric observations, \cite{Scheirich2024a} fit a model in which the motion of the binary was assumed to be Keplerian, allowing for a small quadratic drift in the mean anomaly to account for the combined effects of the binary YORP effect and tides \citep{Scheirich2009,Scheirich2015,Scheirich2021}; they \edit{found} that the total change in the binary's orbital period was $-33.34\pm0.072 (3\sigma)$ m. Most of the orbital period change was nearly instantaneous, but they found that the orbital period dropped by an additional ${\sim}30$ s in the weeks following the impact, with more precise estimated values and associated uncertainties depending on whether the additional orbital period change was assumed to be the result of an exponentially decreasing angular acceleration or an abrupt period change. Using an independent dynamical model fit to radar observations and DART images, in addition to ground-based photometric observations, \cite{Naidu2024a} found a similar result: a total orbital period change of $-33.25\pm 0.03 (1\sigma)$. They also found evidence of an approximately exponential drop in the orbital period of $34\pm15$ s, with an e-folding timescale of $\tau=12\pm7$ d.

It is remarkable that the two solutions for $\Delta P$ lie within $3\sigma$ of one another and that both orbital solutions are consistent with a small, exponential drop in the orbital period in the weeks following the impact. This drop was attributed to a massive cloud of bound ejecta that continued interacting with the binary, providing a drag-like force on Dimorphos and hardening the system \citep{Naidu2024a,Richardson2024a}. The principal concept is angular momentum conservation, according to which some of the binary angular momentum would be transferred to ejecta particles upon future close encounters as Dimorphos \edit{or Didymos  scatter} them out of the system.

While there have been numerous DART ejecta dynamics studies that use more sophisticated numerical models than are presented here, none of them has accounted for the influence of the ejecta back on the binary orbit itself \citep[e.g.,][]{Yu2017,Rossi2022,Moreno2023,Langner2024,Ferrari2025}. Given that the ejecta mass was relatively high and that the orbital period seems to have evolved in the weeks following impact, studying this back reaction of the ejecta on the binary orbit is a key application of an ejecta dynamics model.

\cite{Richardson2024a} demonstrate that this binary hardening effect would be plausible with some order-of-magnitude estimates, which we summarize here. First, they demonstrate that the additional orbital period change should be exponential in time, as the population of bound ejecta particles should decay exponentially. They show that this ejecta clearing timescale should be on the order of days to weeks, which is similar to the e-folding timescale for the binary's orbital period change derived by \cite{Naidu2024a} and \cite{Scheirich2024a}. Second, using angular momentum conservation arguments and some idealized assumptions about ejecta scattering, they demonstrated that the total mass of bound ejecta should be on the order of ${\sim}5\times10^6$ kg, or ${\sim}10^{-3}$ Dimorphos masses. This estimate, although just for the bound component, is consistent with estimates for the ejecta mass derived from telescopic observations \citep[e.g.,][]{Li2023,Graykowski2023,Roth2023,Jewitt2023}. Based on these arguments, \cite{Richardson2024a} conclude that this binary hardening effect is the most likely explanation for the additional orbital period drop, although they acknowledged that direct numerical simulations were needed.

When Dimorphos scatters ejecta out of the system, the binary loses angular momentum and the orbital period decreases, but when ejecta particles re-accrete, they have the opposite effect. Because the DART impact was approximately head-on, most ejecta particles were launched at pericenter on prograde orbits with respect to the binary mutual orbit. Therefore, when they re-impact Didymos or Dimorphos, their angular momenta would, on average, tend to increase the binary angular momentum and orbital period. The evolution of the binary's orbital period, therefore, is a competition between scattering and re-accretion, making it a ripe problem to study with $N$-body methods. 

The simple $N$-body simulations presented in this work will demonstrate that this ``binary hardening'' effect should be impossible for the Didymos system. The order-of-magnitude estimates in the \cite{Richardson2024a} study overlooked the fact that Dimorphos is a very weak scatterer. They assumed that scattering works, then derived the timescale for this to occur and the ejecta mass required to achieve the observed period drop. However, Dimorphos's low mass and close-in orbit make it nearly impossible that ejecta particles would be scattered out of the Didymos system. 

It is useful to consider the Safronov number, which is approximately the ratio of the square of the scatterer's escape speed to the square of its orbital velocity: $\Theta=v_\text{esc}^2/2v_\text{orb}^2$. This is a common parameter in the study of planet formation and is used as an indicator of a planet's scattering properties; when $\Theta$ is greater than unity, close encounters are likely to lead to an ejection, while when $\Theta$ is less than unity, close encounters are likely to result in a collision \citep{Safronov1972,Morbidelli2018}. With an escape speed of ${\sim}9$ cm/s and an orbital velocity of ${\sim}17$ cm/s, Dimorphos has a very small Safronov number ($\Theta{\sim}0.14$), meaning that collisions with nearby bodies should dominate over scattering. This is the same reason why Earth ($\Theta{\sim}0.067$) cannot scatter asteroids out of the Solar System during close encounters but Jupiter ($\Theta{\sim}10.5$) can. As a result of Dimorphos's low mass and close-in orbit, even when it does scatter particles onto wider orbits, they still may not escape the Didymos system and will eventually collide. If Dimorphos were on a much wider orbit, or had an unphysically high density, then ejecta particles could have deeper gravitational encounters and binary hardening would be feasible.

\section{Methods}

We used the REBOUND $N$-body code to model the ejecta and binary dynamics \citep{rebound}. The ejecta particles were treated as gravitationally interacting test particles, meaning that Didymos and Dimorphos gravitationally interact with the test particles but the test particles do not interact with each other. When they collide with either Didymos or Dimorphos, particles are assumed to undergo perfect mergers, conserving mass, volume, and momentum. In addition, any particles traveling beyond Didymos's Hill sphere on an unbound orbit ($e>1$) are deleted from the simulation.

Even for test particles, the default behavior for collisions in REBOUND is to perform a brute force search for particle collisions at each timestep, which is an $\mathcal{O}(n^2)$ problem. Because ejecta-ejecta collisions are expected to be rare, we modified the code to only perform a collision search for Didymos-ejecta and Dimorphos-ejecta collisions, which decreased the complexity to $\mathcal{O}(n)$. We also opted to use REBOUND's second-order leapfrog integrator instead of a more sophisticated option like IAS15. This is because adaptive integrators suffer at early times when many particles are colliding with Dimorphos almost simultaneously, requiring incredibly small timesteps to resolve each collision in time. After a simple convergence test, we opted for an overly conservative fixed timestep of only 6 s, which ensures particles cannot substantially penetrate the surface of either body before a collision is detected and the particle is removed. By default REBOUND only resolves one collision per particle, per timestep, meaning that if multiple particles collide with Dimorphos within one timestep, only one particle will be removed, which can cause unphysical behavior. This behavior in REBOUND is intentional because most users use adaptive integrators that resolve collisions discretely in time. However, in this use case, and especially at early times, many particle collisions can occur within a single timestep, so we made a simple modification to REBOUND's collision routines to allow multiple particles to merge with Didymos or Dimorphos within a single 6 s timestep. 

This study aimed to isolate purely the binary-ejecta interaction, so all higher-order effects are ignored, such as the spin-orbit coupling of the binary, solar tides, solar radiation pressure, and particle-particle gravity and collisions. Including any higher-order effects is unnecessary for our purposes, because none of these things change Dimorphos's escape or orbital speeds, which are the most important determining factors in its scattering efficiency. However, we did perform some tests including solar tides and Didymos's $J_2$ moment using the REBOUNDx library, which had a negligible effect on the binary's orbital period change \citep{reboundx}.

\paragraph{Initial conditions} 

The ejecta cone was directly imaged by LICIAcube and in early \textit{Hubble} Space Telescope images \citep{Li2023,Dotto2024}, which enabled a characterization of the ejecta cone geometry \citep{Deshapriya2023,Hirabayashi2025a}. The ejecta cone is asymmetric, being best described as an elliptical cone having two opening angles: the narrow- and wide-cone opening angles are oriented perpendicular to each other and have respective values of $95\pm6^\circ$ and $133\pm9^\circ$, respectively. The wide-cone opening angle is rotated $28\pm17^\circ$ away from Dimorphos's north-south direction and the cone axis points toward a right ascension and declination of $141\pm4$ and $20\pm8$ in the J2000 frame, respectively. This unique ejecta cone morphology is primarily attributed to the curvature of Dimorphos \citep{Hirabayashi2025a}. In our simulations, Dimorphos is modeled as an idealized sphere, so in order to keep things simple, we ignored the azimuthal asymmetry of the cone. Instead, ejecta positions and velocities are initialized along a symmetric cone with an opening half-angle of $\theta=57^\circ$, which is an average of the wide- and narrow-cone opening angles. We ignored the uncertainty in the cone direction; however, in order to sample some of the uncertainties, and because not all ejecta travel along this perfectly defined cone, we randomly generated velocities and starting positions uniformly within $\pm5^\circ$ of this cone. In addition, we ignored the time dependence of the ejecta formation; all ejecta particles are placed into the simulation at $t=0$, at the surface of Dimorphos, with randomly generated position and velocity vectors along the cone. We used the latest DART SPICE kernels along with the SpiceyPy Python package to help initialize the simulations with the correct geometry \citep{DART_SPICE2023,NAIF,SpiceyPy2020}. The DART SPICE kernels were also used to set the initial binary orbit and body radii.

Due to its closest approach being only ${\sim}3$ min after the DART impact, the images taken by LICIAcube measured ejecta speeds on the order of ${\sim}\text{m/s}$. This study, however, is focused on the slowest ejecta, on the order of ${\sim}\text{cm/s}$, for which there are no direct measurements, as this material would not have been seen until later times. If Dimorphos has near-zero cohesive strength, as geophysical observations of its surface and numerical models of the impact suggest \citep{Barnouin2024a,Robin2024a,Raducan2024b}, then crater-scaling laws and impact models would indicate that a significant fraction of the DART ejecta would have been launched at just a few cm/s \citep{Raducan2024b,Cheng2024}. 

In this study we aim to demonstrate the feasibility of binary hardening, so rather than relying on impact simulations or ejecta scaling laws to generate mass-velocity distributions of the ejecta, we artificially select fixed ejection speeds to test if there are any families of orbits could lead to a contraction in the orbital period. Whether or not sufficient mass would be ejected at these speeds is a secondary question.

\begin{figure}[h!]
   \centering
   \includegraphics[width=\hsize]{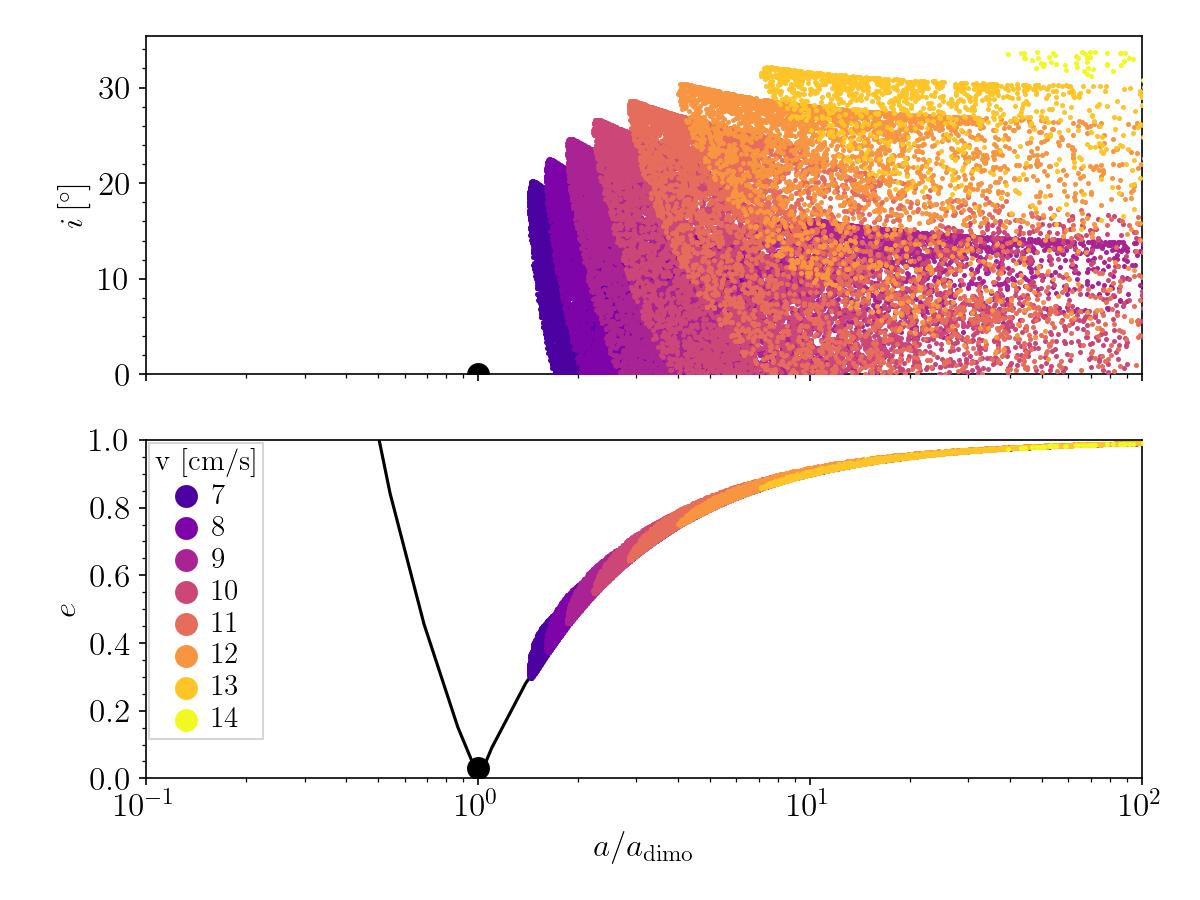}
   \caption{\label{fig:orbElem}Initial orbital elements of ejecta particles. Each ejecta velocity corresponds to a separate simulation. The semimajor axis ($a$) is normalized to the semimajor axis of Dimorphos ($a_\text{dimo}$), and the binary orbital inclination is defined to be 0$^\circ$. The large black dot is the orbit of Dimorphos, and the two black lines are curves of constant apocenter and pericenter that intersect the orbit. We note that these are the instantaneous orbital elements relative to the system barycenter at the beginning of the simulation, when the test particles are still at the surface of Dimorphos. }
\end{figure}

Our core test is a set of eight simulations, each with a single ejecta speed, which varies between 7 and 14 cm/s. In Fig.\ \ref{fig:orbElem}, we show the orbital elements $a,e$, and $i$ of the ejecta particles at the start of each simulation. We can see that higher ejecta speeds translate to higher semimajor axes, eccentricities, and inclinations. Ejecta with speeds below ${\sim}7\text{ cm s}^{-1}$ rapidly re-accrete onto Dimorphos and at speeds above ${\sim}14\text{ cm s}^{-1}$ promptly escape the system, so ejecta at speeds outside of this range cannot contribute to the evolution of the binary's orbital period. Finally, all simulations use $10^4$ equal-mass ejecta particles with a combined mass of 0.01 Dimorphos masses.

\section{Results}

\begin{figure}[h!]
   \centering
   \includegraphics[width=\hsize]{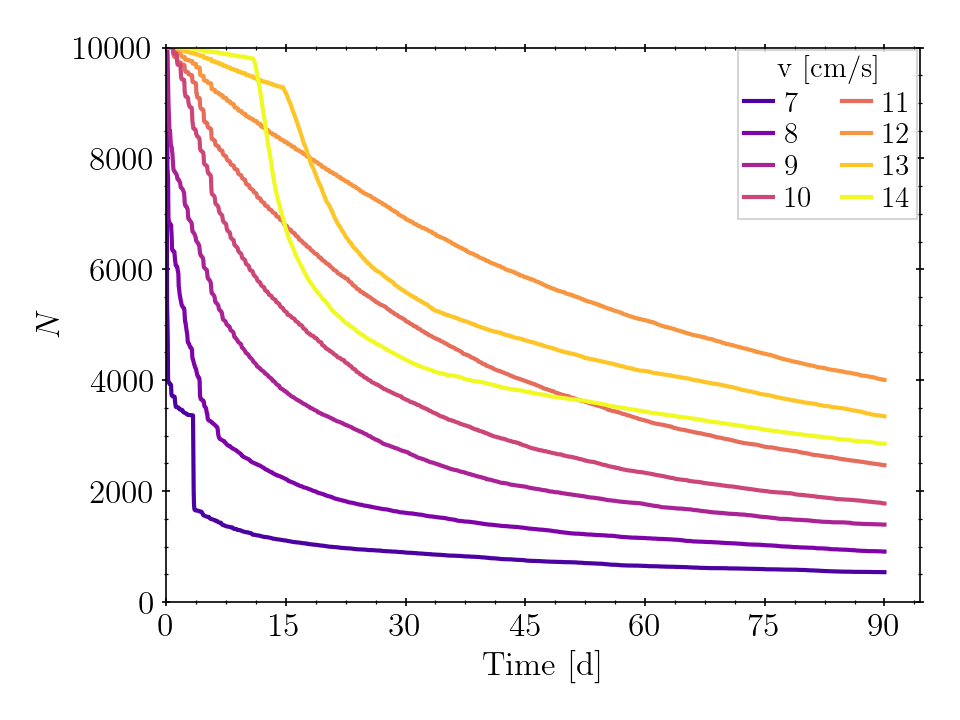}
   \caption{\label{fig:N_vs_t}Number of ejecta particles over time for each simulation, which decays exponentially as expected. The sudden drop in $N$ for the $7 \text{ cm s}^{-1}$ case at $t{\sim}5\text{ d}$ is due to the accretion of a large group of particles {(see movie)}. The kinks around ${\sim}15$ d for the 13 and 14 cm s$^{-1}$ cases are due to particles reaching the Hill sphere and being removed from the simulation. 
   }
\end{figure}

First, we show the remaining number of ejecta particles in the system for each simulation in Fig.\ \ref{fig:N_vs_t}. The ejecta population clearly decays exponentially, as predicted by \cite{Richardson2024a} and demonstrated by other ejecta dynamics models \citep[e.g.,][]{Langner2024}. 

\begin{figure}[h!]
   \centering
   \includegraphics[width=\hsize]{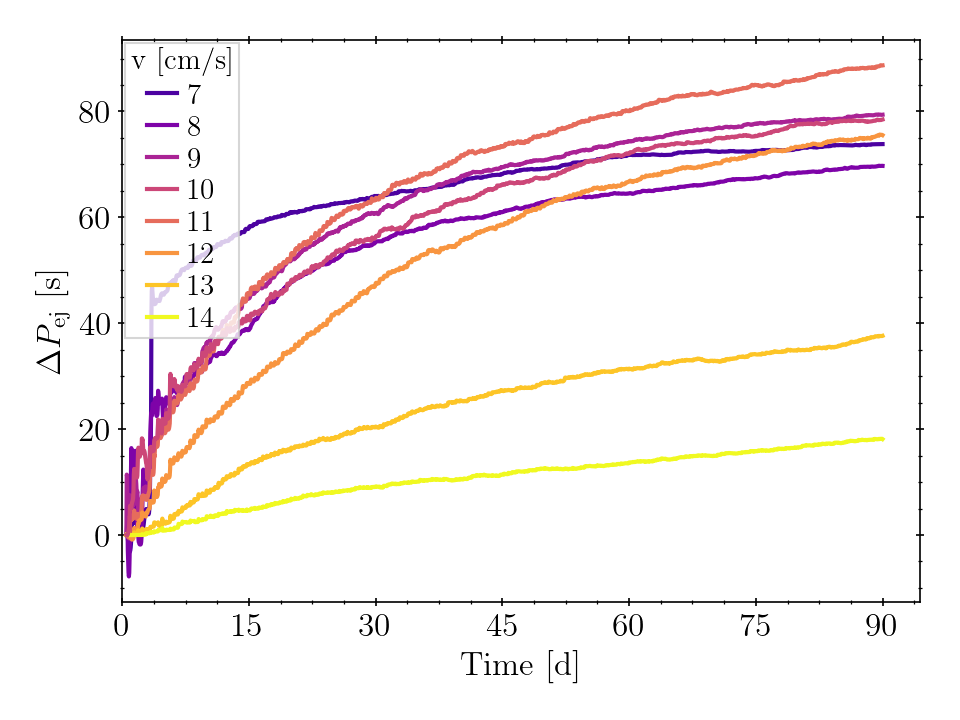}\caption{\label{fig:dP_vs_t}Change in the binary's orbital period due to the interaction with ejecta, as a function of time.   }
\end{figure}

In Fig.\ \ref{fig:dP_vs_t} we show our main result, that all tested ejecta speeds lead to an increase in the binary's orbital period. At higher speeds, the effect is less pronounced because most ejecta immediately escapes the system with little interaction with the  binary. We also see that the change in period is logarithmic in time, for the same reasons predicted in \cite{Richardson2024a}, although the behavior is in the opposite direction with the period increasing rather than decreasing. \edit{In our simulations, the binary's orbital period is based on Dimorphos's Keplerian orbit in Jacobi coordinates. In other words, Dimorphos's instantaneous semimajor axis is computed relative to the system barycenter at each point in time, and then the corresponding orbital period is computed according to Kepler's 3rd Law. }  If spin-orbit coupling were accounted for, there would be additional fluctuations on the order of ${\sim}10$s of seconds in the binary's orbital period due to angular momentum exchange between Dimorphos's spin and the mutual orbit \citep{Meyer2021b}.

Here, $\Delta P$ is the change relative to the orbital period 12 hours after impact: $\Delta P = P(t) - P(t=12 \text{ h})$. The choice of 12 h is somewhat arbitrary but was chosen because it is approximately one binary orbital period. This choice was made so that $\Delta P$ shows only the effect of binary-ejecta interactions, with the immediate ${\sim}30$ m orbital period change already accounted for. This also masks any transient behavior in simulations with low ejecta speeds, where a significant fraction of ejecta is re-accreted within the first few hours. Although this transient behavior may have occurred in reality, it would not have been observed, as the evolution of the orbit was determined by mutual event timings in the weeks and months following the impact. We include movies for each of the eight simulations, which show the evolution of the ejecta cloud. Due to the large file sizes, the movies are stored on a separate Zenodo repository\footnote{The Zenodo repository with accompanying movies can be found at the following DOI/URL: \url{https://doi.org/10.5281/zenodo.15575491}}

\paragraph{What would it take for binary hardening to work?}
As a sanity check, and to demonstrate that binary hardening would have occurred under different circumstances, we tested the conditions that would allow Dimorphos's orbital period to decrease due to a massive ejecta cloud. 

We performed simulations similar to those shown above with a fixed ejecta velocity, except we also varied Dimorphos's orbital semimajor axis. As the binary separation increases, Dimorphos's orbital velocity decreases, which effectively makes scattering more efficient relative to accretion. To simplify things, we placed Dimorphos on a circular orbit at a specified semimajor axis and pointed the eject cone exactly in the direction of Dimorphos's orbital velocity, whereas in the previous section, Dimorphos was initialized with a nonzero eccentricity ($e{\sim}0.03$) and the ejecta cone was slightly misaligned with respect to the instantaneous orbital velocity direction. However, all other parameters are kept the same, such as the ejecta cone opening angle, and the number and mass of ejecta particles.

\begin{figure}[h!]
   \centering
   \includegraphics[width=\hsize]{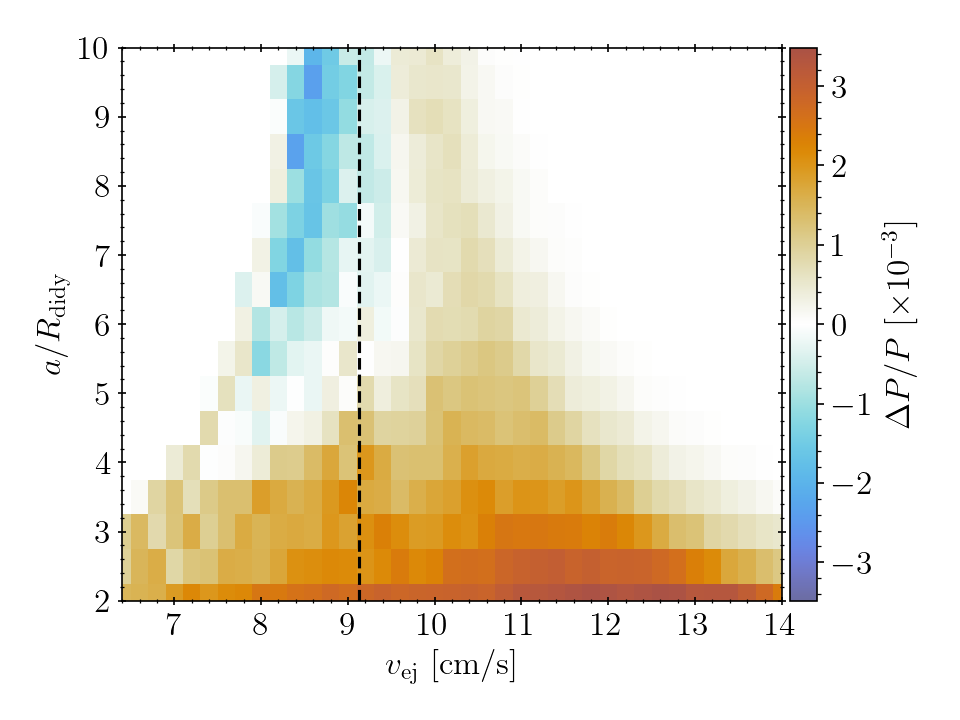}
   \caption{\label{fig:dP_grid}Fractional change in the binary's orbital period as a function of the ejecta velocity and Dimorphos's semimajor axis ($a$) normalized to Didymos's radius ($R_\text{didy}$). The dotted black line is the surface escape speed of Dimorphos, and Dimorphos's actual semimajor axis is ${\sim}3R_\text{didy}$.}
\end{figure}

In Fig.\ \ref{fig:dP_grid} we plot the fractional change in binary period as a function of the ejecta speed and Dimorphos's semimajor axis normalized to Didymos's radius, $R_\text{didy}$. Recall that Dimorphos's real semimajor axis is $a/R_\text{didy}{\sim}3$. At this distance, simulations at all ejecta speeds lead to a net increase in the binary's orbital period. As the binary separation increases, Dimorphos's orbital velocity decreases (i.e., the Safronov number increases), increasing the efficiency of scattering until eventually, near $a/R_\text{didy}{\sim}5$, cases with ejecta speeds near Dimorphos's escape speed undergo binary hardening and the orbital period decreases. 

With increasing binary separation, we also see that ejecta speeds on both the low and high end lead to negligible orbital period change. For higher ejecta speeds ($v_\text{ej}\gtrsim10\text{ cm s}^{-1}$), particles are immediately launched on escape trajectories and do not interact strongly with the binary. On the lower end ($v_\text{ej}\lesssim 8\text{ cm s}^{-1}$), below the escape speed, meaning that they cannot reach the Hill sphere and are re-accreted within a single orbital period. In between these end-member cases, there is some dependence on the ejection speed. Particles ejected just below the escape speed, around ${\sim} 8.5 \text{ cm s}^{-1}$ seem to be the most effective at hardening the binary. We attribute this to the relatively low angular momentum of these orbits: these particles have just enough energy to escape Dimorphos and orbit Didymos. If they re-accrete on Dimorphos, they only marginally increase the orbital period because their angular momentum is relatively low. When Dimorphos scatters them out of the system, however, it must transfer a significant amount of angular momentum to remove them, leading to a larger drop in the period. Then for higher ejecta speeds, just above the escape speed, we have the opposite case; these particles already have (relatively) high angular momenta, so when they are ejected, they negligibly decrease the binary's orbital period because there is very little angular momentum left to extract. However, when they re-accrete, they transfer their large angular momentum back to the binary, leading to a net increase in the binary's orbital period. 

These results demonstrate that, for the Didymos system as we know it, binary hardening is a very unlikely explanation for the apparent orbital period drop. Dimorphos would need to have a much wider semimajor axis of $a\gtrsim5R_\text{didy}$ in order for low speed ejecta to have a chance at shrinking the binary's orbital period. 

\section{Discussion}
The intention of this study was to verify the claims of \cite{Richardson2024a} and \cite{Naidu2024a}, and to explore what ejecta mass-velocity distributions might decrease the binary's orbital period with the correct magnitude and timescale. This could then be used to constrain the material properties of Dimorphos itself, because the ejecta mass-velocity distribution is closely related to the cohesive strength \citep{Housen2011}. Instead, we reached the opposite conclusion for the simple reason that Dimorphos is too weak of a scatterer to eject sufficient angular momentum from the system. Therefore, we argue that binary hardening is extremely unlikely to have occurred for the Didymos system following the DART impact. In our model, all ejecta speed leads to a net increase in the binary's orbital period, meaning that this result is also independent of the ejecta mass-velocity distribution. 

Dimorphos's escape speed is small compared to its orbital velocity, meaning that it has a low Safronov number, and is therefore a weak scatterer. Effectively, the re-accretion of ejecta particles adds back more angular momentum to the binary than Dimorphos is capable of removing by gravitational scattering. If Dimorphos were a more powerful scatterer --- a scenario that would require Dimorphos to have either a much higher escape velocity or a much lower orbital speed, both of which are unphysical --- then binary hardening could explain the anomalous orbital period drop.  

It is plausible that un-modeled higher-order effects, such as self-gravity within the ejecta cone or solar radiation pressure, could put particles on orbits that can avoid collisions and be preferentially scattered. However, we think that a higher-order model would reach the same conclusion because adding additional physics does not change the fundamental fact that Dimorphos's escape speed is much less than its orbital velocity, making it a poor scatterer. We also note that some basic tests were performed accounting for solar tides and the primary's $J_2$, both of which were determined have a negligible effect. A recent study, which determined the change in the Didymos system's heliocentric orbit, suggests that Dimorphos is under-dense relative to Didymos \citep[{Makadia, pers. comm.,}][]{Makadia2024}. This would further decrease Dimorphos's escape speed, which would make binary hardening even less viable.

Given that the uncertainties in the apparent ${\sim} 30$ s orbital period drop were similar in magnitude to the change itself, this detection is somewhat marginal, meaning that is plausible that the orbital period did not evolve in the weeks following the DART impact \citep{Scheirich2024a,Naidu2024a}. However, additional mutual event data taken in early 2025 yield an improved post-impact orbital period of $P_\text{orb}=11.3667 \pm 0.0002 (3\sigma)$ h, which is ${\sim}3$ s shorter (and with smaller uncertainties) than the original post-impact orbital period published by \cite{Scheirich2024a} and \cite{Knight2025_epsc}. This does not confirm whether the original ${\sim}30$ s truly occurred in the first several weeks after impact, but it does not rule it out either. It at least suggests that the orbital period evolved between the DART impact and early 2025. 

If we assume the ${\sim}30$ s drop was real, then this presents an interesting mystery. Models of the DART impact and the ejecta morphology both suggest that a significant fraction of the ejecta mass was launched at speeds comparable to Dimorphos's escape speed \citep{Raducan2024b,Cheng2024,Ferrari2025}. This study shows that most of this material would have re-accreted onto Didymos or Dimorphos in weeks following impact, leading to a slight increase in the orbital period, with the magnitude and timescale depending on the precise mass-velocity distribution of ejecta. If the anomalous orbital period drop is real, then some additional mechanism must have overcome the effect of the ejecta accretion, and led to a net decrease in the orbital period. 

We leave the investigation of alternative orbital period-reducing mechanisms to future work, but discuss some possibilities here. One way to reduce the binary's orbital period is to increase the oblateness  ($J_2$) of Didymos. Prior to the DART impact, it was speculated that Didymos could undergo structural failure and global reshaping, which would decrease the binary's orbital period \citep[e.g.,][]{Hirabayashi2019c,Hirabayashi2022}. However, to reduce the binary's orbital period by some tens of seconds, as would be required here, the Didymos's global shape change would be so large that its spin period would also increase by some tens  of seconds to conserve angular momentum \citep{Nakano2022}. No such spin period change was detected for Didymos, so we rule out global reshaping of Didymos as an explanation for Dimorphos's orbital period drop \citep{Pravec2024a}. 

Although Dimorphos only contains ${\sim}1\%$ of the system's mass, its own reshaping, and the associated change in the mutual gravitational potential, could also sufficiently reduce the orbital period \citep{Nakano2022}. Indeed, \cite{Nakano2024a} demonstrated that the impact-induced global reshaping of Dimorphos predicted by \cite{Raducan2024b} may have been responsible for tens or hundreds of seconds of the binary's orbital period change. This global reshaping and associated orbital period change would have occurred immediately following the impact and is separate from the additional orbital period change considered here. We speculate that Dimorphos's additional orbital period drop in the weeks and months following the impact could be due to additional reshaping that occurred either continually or instantaneously at some epoch in the months following impact. We argue that the most likely cause of Dimorphos's continued reshaping could be a result of its chaotic rotation, which was likely triggered by the DART impact \citep{Agrusa2021}. Both the post-impact Dimorphos light curves and a sudden drop in the binary eccentricity about ${\sim}70$ d after impact are consistent with the onset of non-principal axis (NPA) rotation \citep{Pravec2024a,Scheirich2024a,Meyer2023b}. It was also predicted that NPA rotation could be sufficiently violent to trigger landslides or surface motion on Dimorphos \citep{Agrusa2022a,Agrusa2022b}, although these studies relied on assumptions about the pre-impact shape of Dimorphos. Given that it appears Dimorphos has already reshaped, that it underwent NPA rotation, and that NPA rotation can trigger surface motion, we suggest that a rotation-induced reshaping of Dimorphos is a plausible explanation for the anomalous drop in the binary's orbital period. However, we caution that additional studies are needed to reach a robust conclusion.

\section{Conclusions}

Since Dimorphos is a weak scatterer, we find that the gravitational interaction of bound ejecta is an unlikely explanation for the additional drop in Dimorphos's orbital period in the weeks following the DART impact. In fact, Dimorphos's orbital period likely increased as a result of this process since collisions, which tend to add angular momentum back to the binary, dominate over scattering events, which tend to extract angular momentum from the binary. 

If the orbital period did drop by tens of seconds in the months following the DART impact, then some additional mechanism is needed to explain this process. We suggest that an additional reshaping of Dimorphos, due to its excited post-impact rotation state, could provide an explanation. However, we leave a more detailed investigation of possible mechanisms to future work.

\begin{acknowledgements}
   We thank Petr Pravec, Petr Scheirich, and Patrick Michel for useful discussions. \edit{We thank the anonymous referee for their careful examination of this manuscript.} H.A. was supported by the French government, through the UCA J.E.D.I. Investments in the Future project managed by the National Research Agency (ANR) with the reference number ANR-15-IDEX-01.
\end{acknowledgements}

\bibliographystyle{aa} 
\bibliography{references} 

\end{document}